# Topological and superconducting properties of two-dimensional $C_{6-2x}(BN)_x$ biphenylene network: a first-principles investigation

Guang F. Yang,[1] Hong X. Song,[1,*] Dan Wang,[1] Hao Wang,[1] and Hua Y. Geng[1,2,*]

[1]*National Key Laboratory of Shock Wave and Detonation Physics, Institute of Fluid Physics, China Academy of Engineering Physics, Mianyang, Sichuan 621900, P. R. China;*

[2]*HEDPS, Center for Applied Physics and Technology, and College of Engineering, Peking University, Beijing 100871, P. R. China;*

## Abstract

First-principles calculations have been used to investigate the electronic and topological properties of the two-dimensional $C_{6-2x}(BN)_x$ biphenylene network, a graphene-like structure composed of not only hexagonal ring but also octagonal and square rings. Nontrivial topological properties have been found in two of them, with a stoichiometry of $C_4BN$ and $C_2(BN)_2$. The former $C_4BN$ is predicted to be a type-II Dirac semimetal with a superconducting critical temperature Tc=0.38K, which is similar to the pure carbon biphenylene network (C-BPN). The latter shows a novel isolated edge state exists between the conduction and valence bands. By regulation of strains and virtual-crystal approximation calculations, we found the annihilation of two pairs of Dirac points (DPs) in the non-high symmetric region (non-HSR) causes the two corresponding edge states stick together to generate this isolated edge state. In addition, we found that one pair of DPs arises from the shift of DPs in the C-BPN, while another new pair of DPs emerges around the Time Reversal Invariant Momenta (TRIM) point X due to the doping of boron and nitrogen. We constructed a tight-binding (TB) model to reveal the mechanism of forming the isolated edge state from the C-BPN to $C_2(BN)_2$. This study not only demonstrates the existence and mechanism of forming the isolated edge state in semimetals, but also provides an example in which the DPs can move away from the high-symmetry region.

* Corresponding authors:
E-mail address: s102genghy@caep.cn (Hua Y. Geng), hxsong555@163.com (Hong X. Song).

## I. Introduction

Shortly after the experimental fabrication of the two-dimensional (2D) graphene through the mechanical stripping method[1], its unique electronic[2-4], thermal[5-8], and mechanical[9-10] properties were systematically reported. Its honeycomb lattice also inspired the introduction of many famous models in topological band theory, including the Haldane Model[11] and the Kane-Mele Model[12]. The predicted Dirac cones at the Fermi level ($E_F$) in graphene play a central role in understanding the topologically protected quantum states in topological insulators and Dirac semimetals[11-16]. Recently, Fan *et al.* successfully synthesized biphenylene[17], another carbon allotrope first predicted in 1997[18]. It consists of adjacent octagonal, hexagonal, and square rings (4-6-8 rings), which is commonly referred to as the "ohs" structure. Despite only being synthesized two years ago, the ohs structure has already attracted a significant amount of interest due to its novel properties in electronic band structure[19-21], phonon heat transport[22-24], thermoelectricity[25], mechanical properties[23], hydrogen evolution reaction[26], and superconductivity(Tc = 0.59K)[19,27]. Moreover, the above mentioned properties of the biphenylene network (BPN) could be further regulated by atomic doping or adsorption[28-38].

In particular, electronic structure calculations revealed that there is a pair of tilted type-II Dirac cones along the $\Gamma$-Y direction of the Brillouin zone and an edge state connecting the Dirac cones along the armchair edge, indicating it is a 2D Dirac semimetal[19]. Further investigations have indicated that the superconducting critical temperature (Tc) of biphenylene can be increased from 0.59K to 3.91K by absorbing Li atom on the hexagonal rings of BPN. Under the application of biaxial tensile strain, Tc can be further enhanced up to 15.86K[33]. Given those graphene-like materials with honeycomb lattice have been extensively studied and have proven to be a fruitful testing ground for exploring Dirac materials and topological concepts, we can reasonably anticipate that the BPN structure composed of 4-6-8 rings may also share similar features in electronic and topological properties. Therefore, expanding the BPN system beyond pure carbon is valuable to unveil the physics of this type of material.

Considering that carbon can form strong covalent bonds with boron and nitrogen because they are similar in atomic size and electronegativity[39-44]. It is possible to substitute part of the carbon-carbon bonds in biphenylene with boron-nitrogen bonds, similar to graphene-like nanomembranes composed of boron, carbon, and nitrogen. Recently, Lv *et al.* have discovered two types of $C_{6-2x}(BN)_x$ structures[34] by using the first-principles structural search. By controlling the shape of the band structure, they were able to achieve an improvement in the thermoelectric properties. The above-mentioned studies have inspired us to investigate the electronic, topological and superconducting properties of the 2D $C_{6-2x}(BN)_x$ ohs structure.

In this work, we constructed two configurations of $C_4BN$ and $C_2(BN)_2$, their dynamic stability is examined by phonon dispersion calculation. For $C_4BN$, the energy band structure is similar to pure carbon BPN(C-BPN) with a pair of type-II Dirac points along the $\Gamma$-Y line [19]. For $C_2(BN)_2$, we found that an isolated edge state exists between the conduction and valence band. By the regulation of strains, the conduction and valence bands overlap in reverse to produce two pairs of Dirac points (DPs) in the non-high symmetric region (non-HSR). The isolated edge state is split into two edge states. With the shift of the DPs in the Brillouin zone, a pair of DPs with opposite chirality finally annihilates at the Time Reversal Invariant Momenta (TRIM), and the corresponding edge state disappears. By the virtual-crystal approximation calculations and tight-binding (TB) model analysis, we illustrated the mechanism of forming this isolated edge state. For superconductivity, we found that the electron-phonon coupling (EPC) strength in $C_4BN$ is weakened compared to C-BPN, with a critical temperature Tc=0.38 K, due to the doping of boron and nitrogen.

## II. Computational methods

The first-principles calculations were performed within the framework of density functional theory[45-46] (DFT) implemented in the VASP[47] package. We used the Perdew-Burke-Eznerhof (PBE) parameterization of generalized-gradient approximation (GGA) to describe the exchange-correlation energy[48]. The projector-augmented-wave (PAW)[49] was chosen to treat the electron-ion interaction. The spin-orbit coupling (SOC) effect was considered in the calculation of electronic properties. In the structural optimization, all atoms were fully relaxed until the Hellmann-Feynman forces on each atom are smaller than 0.0001eV/Å. A kinetic energy cutoff of 550 eV was adopted with a Gaussian width of 0.1 eV for the smearing. A $14 \times 12 \times 1$ Monkhorst-Pack k-points mesh[50] was employed to sample the Brillouin zone. To evaluate the structural stability, we conducted *ab-initio* molecular dynamics simulations using a $5 \times 5 \times 1$ supercell with only Γ-point included.

To investigate the electronic structure of nanoribbons and edge states, a TB

Hamiltonian was constructed by the method of maximally localized Wannier functions (MLWFs)[51] by using the *WANNIER90* package[52-53]. The nontrivial edged states are calculated from the imaginary part of surface Green's function by using *WANNIERTOOLS* package[54].

For the superconducting calculation, the phonon dispersion and electron-phonon coupling calculations were performed based on density functional perturbation theory (DFPT) with norm-conserving pseudopotentials [55] as implemented in the Phonon module of the QUANTUM ESPRESSO (QE) package [56] with a $10 \times 8 \times 1$ q-mesh and $40 \times 32 \times 1$ k-point grid. The cutoff for kinetic energy and charge density are 80 and 400 Ry, respectively. The electron phonon coupling (EPC) $\lambda_{q\nu}$ with mode $\nu$ and wave vector $\boldsymbol{q}$ was calculated as[57]:

$$\lambda_{q\nu} = \frac{\gamma_{q\nu}}{\pi h N(E_F)\omega_{q\nu}^2},\qquad(1)$$

where $\omega_{q\nu}$ represents the phonon frequency, and $N(E_F)$ is the electronic density of states (DOS) near $E_F$. The phonon linewidth $\gamma_{q\nu}$ reads:

$$\gamma_{q\nu} = \frac{2\pi\omega_{q\nu}}{\Omega_{\text{BZ}}} \sum_{k,n,m} \left|g_{kn,k+qm}^{\nu}\right|^2 \delta(\varepsilon_{kn} - E_F)\,\delta(\varepsilon_{k+qm} - E_F),\qquad(2)$$

where $\Omega_{\text{BZ}}$ is the volume of Brillouin zone, $\varepsilon_{kn}(\varepsilon_{k+qm})$ represent the eigenvalues of Kohn-Sham Hamiltonian, and $g_{kn,k+qm}^{\nu}$ is the elements of the electron-phonon matrix [58], respectively. The Eliashberg electron-phonon spectral function $\alpha^2 F(\omega)$ ,which describes the coupling between the phonon and the electrons near the Fermi level, is calculated as follow:

$$\alpha^2 F(\omega) = \frac{1}{2\pi N(E_F)} \sum_{q\nu} \frac{\gamma_{q\nu}}{\omega_{q\nu}} \delta(\omega - \omega_{q\nu}).\qquad(3)$$

The total EPC $\lambda(\omega)$ can be calculated by integrating the Eliashberg spectral function $\alpha^2 F(\omega)$[59]:

$$\lambda(\omega) = \sum_{q\nu} \lambda_{q\nu} = 2 \int_0^\omega \frac{\alpha^2 F(\omega')}{\omega'} d\omega'. \tag{4}$$

The superconducting critical temperature $T_c$ can be determined using the McMillan-Allen-Dynes formula [58]

$$T_c = f_1 f_2 \frac{\omega_{log}}{1.2} \exp\left[-\frac{1.04(1+\lambda)}{\lambda - \mu^*(1+0.62\lambda)}\right], \tag{5}$$

where $\mu^* = 0.1$ is the effective screened Coulomb repulsion constant, $\omega_{log}$ represents the logarithmic average frequency,

$$\omega_{log} = \exp\left[\frac{2}{\lambda} \int_0^\infty \frac{d\omega}{\omega} \alpha^2 F(\omega) \log \omega\right], \tag{6}$$

## III. Results and discussion

## A. Electronic and topological properties

We constructed two different configurations of $C_{6-2x}(BN)_x$ monolayer by replacing the C-C bonds in the ohs structure with B-N bonds, the structural parameters are summarized in Table S1 of the supplementary materials (SM). We only focus on the $C_2(BN)_2$ configuration in the following discussion since the $C_4BN$ has similar properties [see the SM] with C-BPN. In terms of dynamics, they are all stable, which can be confirmed by the phonon spectrum calculation, as shown in Fig. 1(c). Although a small portion of imaginary frequencies exist around the $\Gamma$ point in Fig. 1(c), it's a common and negligible phenomenon in 2D materials. The unit cell of $C_2(BN)_2$ with a rectangular lattice structure [see Fig. 1(a)], belongs to the $P2/m$ space group with a centrosymmetry. Fig. 1(a) shows the electron localization function (ELF) along the [001] plane between ELF =0 and 0.9, where the value of 0.9 corresponds to a perfect localization. There is a noticeable electron localization in the middle of each bond, which confirms the $C_2(BN)_2$ monolayer is a covalent compound.

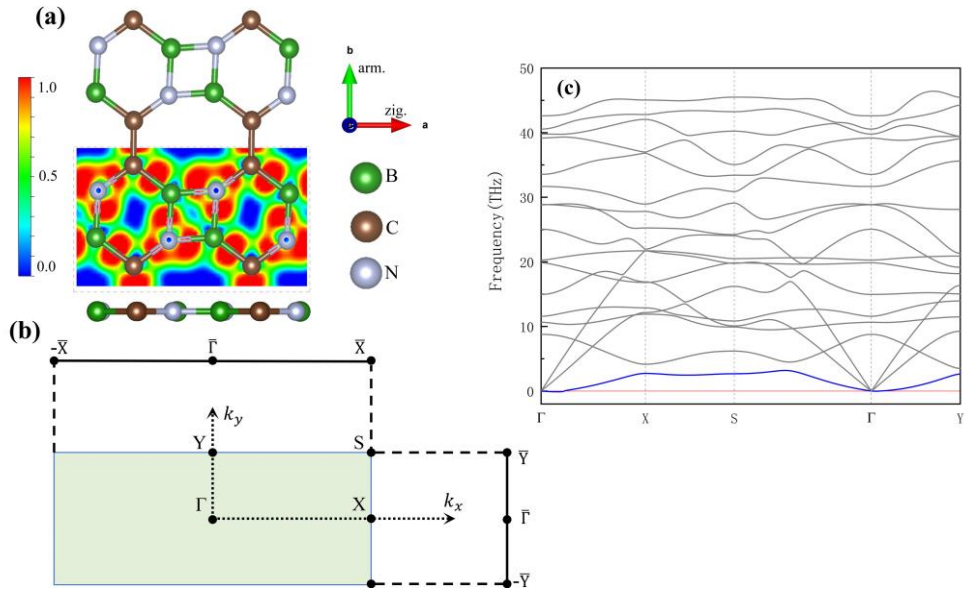

**Fig 1.** The atomic structure and electron localization function of $C_2(BN)_2$(a). Brillouin zone together with the high-symmetry k-points for $C_2(BN)_2$(b). Phonon spectrum(c) of $C_2(BN)_2$.

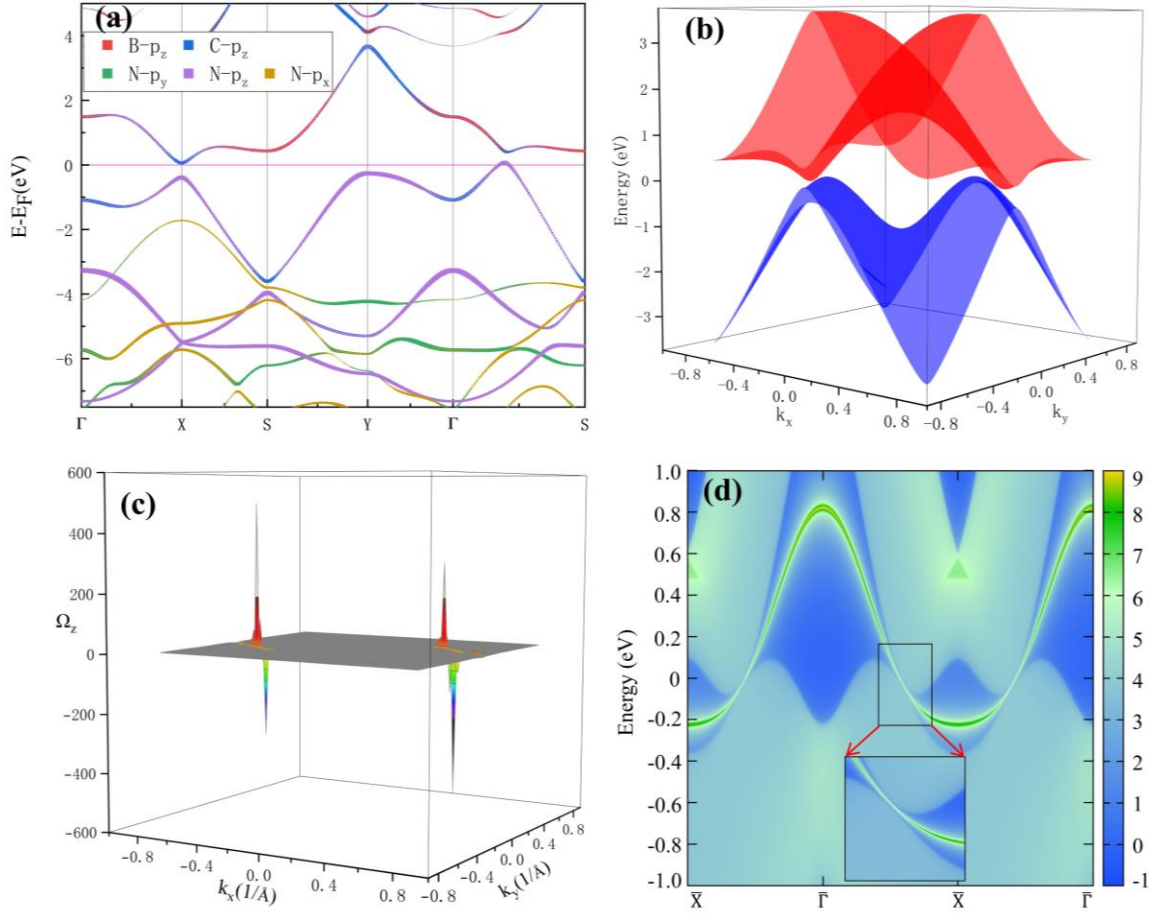

**Fig 2.** Orbital-resolved bulk band structure with SOC (a), electronic band structure for the conduction band and valence band (b) and Berry curvature $\Omega_z$ (c) in the 2D Brillouin zone of the $C_2(BN)_2$. Edge DOS(d) along the zigzag cut [see Fig. 1(a)] of the $C_2(BN)_2$.

Fig. 2(a) shows the calculated orbital-resolved bulk band structures along the high symmetry lines in the Brillouin zone. The band structure indicates that $C_2(BN)_2$ is metallic as the valence band crosses the Fermi level. We calculated the bulk band structure and found that the local energy gap exists all over the Brillouin zone [Fig. 2(b)]. We calculated the Berry curvature using the following formula: $\boldsymbol{\Omega}_n(\boldsymbol{k}) = \nabla_{\boldsymbol{k}} \times \boldsymbol{A}_n(\boldsymbol{k}) = \nabla_{\boldsymbol{k}} \times i\langle u_n(\boldsymbol{k})|\nabla_{\boldsymbol{k}}|u_n(\boldsymbol{k})\rangle$, where $\boldsymbol{A}_n(\boldsymbol{k})$ is the Berry connection that is

similar to the vector potential in a magnetic field and $|u_n(\boldsymbol{k})\rangle$ represents the Bloch wave functions labeled by the crystal momentum $\boldsymbol{k}$ and band index $n$. The result is illustrated in Fig. 2(c), nonzero Berry curvature exists in certain regions. The corresponding Berry phase is given by $\gamma_n = \oint_c \boldsymbol{A}_n(\boldsymbol{k}) \cdot d\boldsymbol{k} = \iint_s \boldsymbol{\Omega}_n(\boldsymbol{k}) \cdot d\boldsymbol{S}$. Where $S$ defines an area whose boundary $c$ encloses the nonzero Berry curvature. The calculated Berry phase is $\pm\pi$. To confirm their edge states, we constructed a tight-binding Hamiltonian by the *WANNIER90* package. The surface electronic structure of a semi-infinite monolayer of $C_2(BN)_2$ was then calculated by using Green's function method. The computed edge state along the zigzag-edged ribbon is shown in Fig. 2(d). An isolated complete edge state appears between the conduction and valence bands.

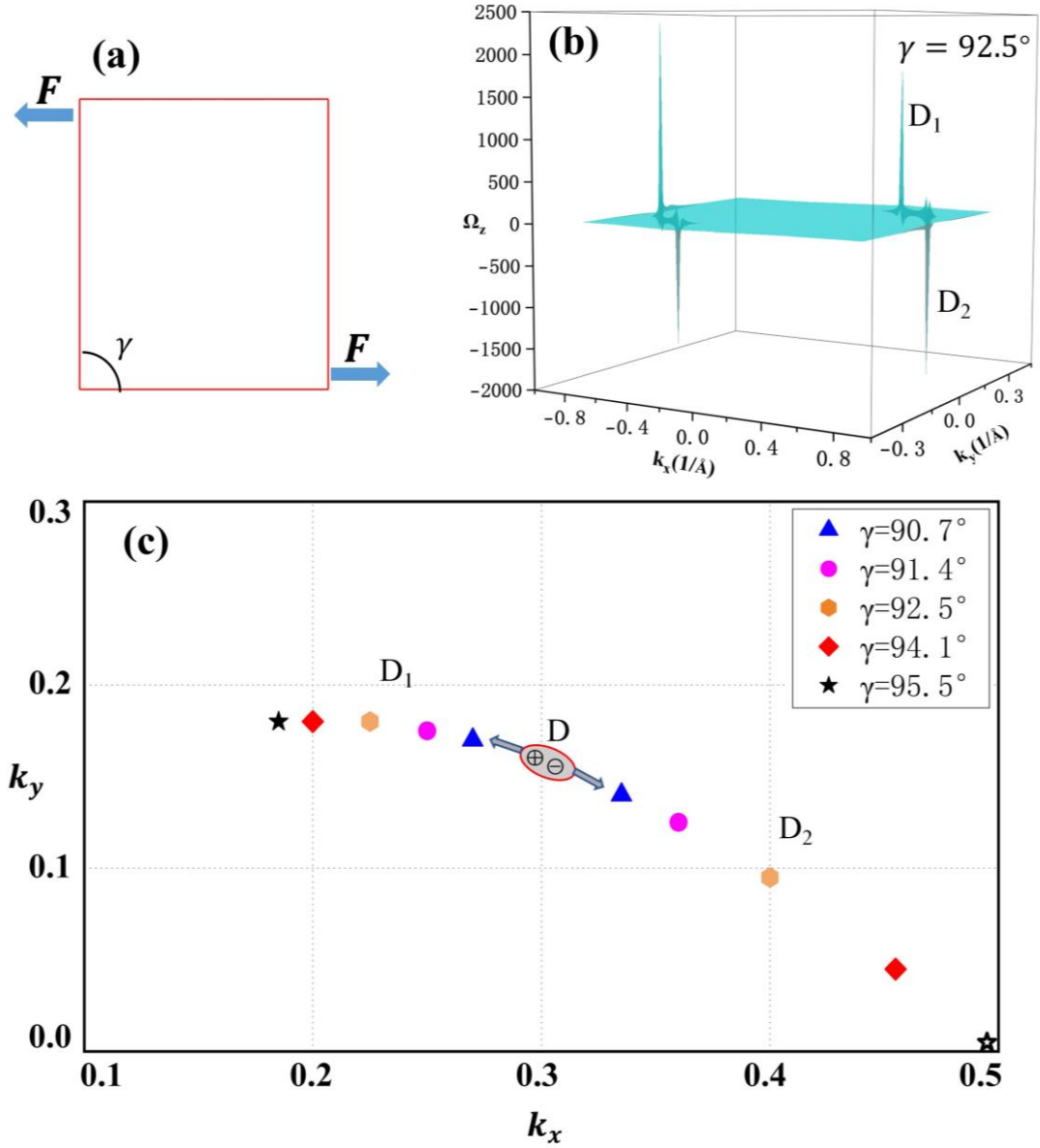

**Fig 3.** The schematic diagram of the strain regulations by increasing cell angle $\gamma$ from 90° to 95.5° (a); the Berry curvature $\Omega_z$ at the $\gamma = 92.5°$ (b); the evolution of the location of the Dirac points with the positive chirality ($D_1$) and negative chirality ($D_2$), in the first quadrant of Brillouin zone (c). The open star represents the annihilation of a paired of Dirac points with the opposite chirality located at $D_2$.

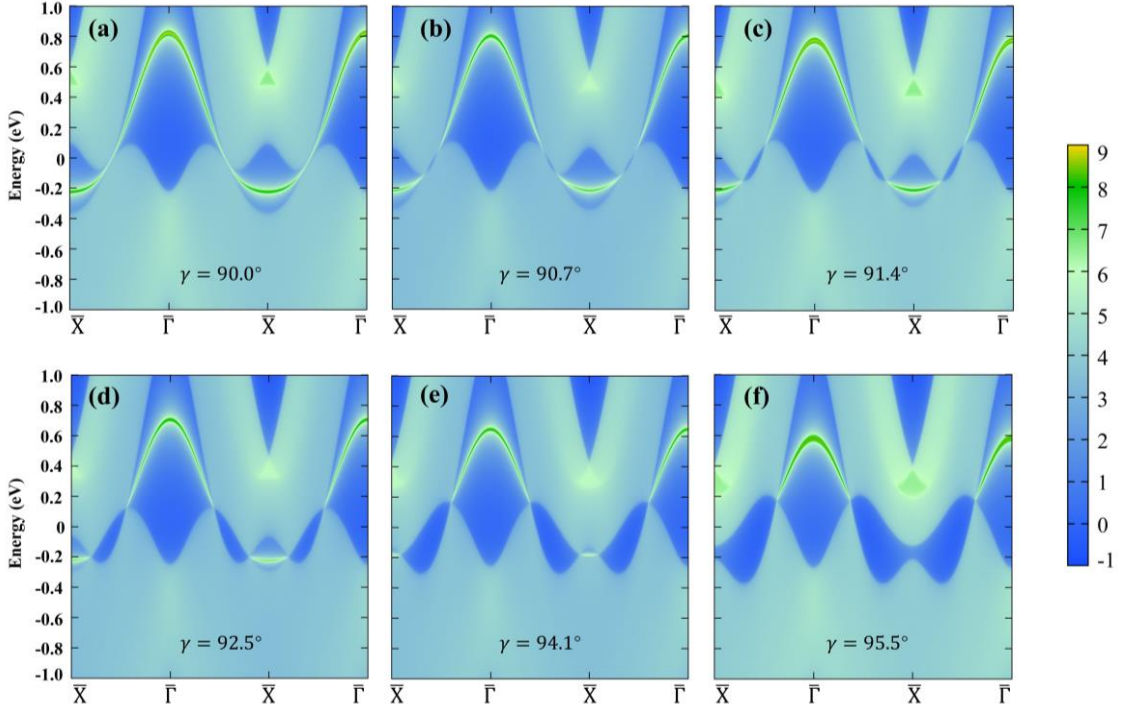

**Fig 4.** The edge DOS along the zigzag cut of the $C_2(BN)_2$ with different cell angle $\gamma$.

To further understand the isolated edge state and the nonzero local Berry curvature for the $C_2(BN)_2$, we increased the cell angle $\gamma$ by the regulation of strains as shown in Fig. 3(a) while keeping the lattice length constant. In the whole process, the position of the Dirac points with positive chirality ($D_1$) and negative chirality ($D_2$) is shown in Fig. 3(c), and the Dirac points with opposite chirality are located symmetrically about the origin. The gray oval region represents the region of the initial structure ($\gamma = 90°$) with the Berry phase of $\pm\pi$. Conveniently, we represent position D on the center of Dirac points at the $\gamma = 90.7°$ as the gray oval region [Fig. 3(c)]. With increasing $\gamma$, two Dirac points ($D_1$ and $D_2$) appear on each side of the oval region. Without loss of generality, we only show the Berry curvature [Fig. 3(b)] at the $\gamma = 92.5°$ and all others could be found in Fig. S1 of the SM. Figure 4 shows the corresponding edge states under the strain regulation. We found that the creation of two pairs of Dirac points causes the "isolated complete edge state" [Fig. 4(a)] is split into two parts of edge states connecting each pair of Dirac points. With the $\gamma = 95.5°$, a pair of Dirac points(D2) annihilates at the TRIM point X [the open star shown in Fig. 3(c)] and the corresponding edge state disappears [Fig. 4(f)], resulting a topological transition.

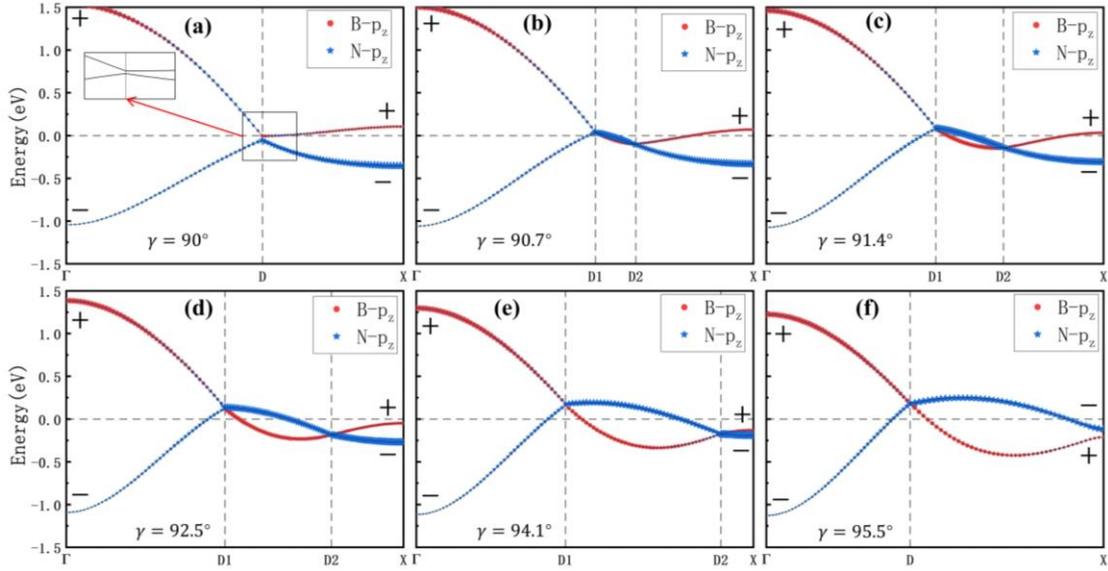

**Fig 5.** Orbital-resolved bulk band structure around the Dirac points in the process of strain regulation. The sign "＋" and "－" represent the parity eigenvalue of the conduction band and the valence band at the two TRIM points $\Gamma$ and X.

To study what happens when the Dirac points emerge and annihilate, we calculated the orbital-resolved band structure around these Dirac points in the whole process of strain regulation. The corresponding 3D energy band structures are shown in Fig. S2 of the SM. As shown in Fig. 5(a-c), the emergency of two pairs of Dirac points is accompanied by an overlap of energy bands. In the process of annihilation, as shown in Fig. 5(d-f), the annihilation of Dirac points(D2) causes the parity change at the point

X. We also calculated the $\mathbb{Z}_2$ topological invariant to confirm its topological nature. This number comes directly from the parity eigenvalue of the occupied bands at the TRIM points. There parities $\xi(i)$ of the 12 occupied valence bands for the $C_2(BN)_2$ monolayer are summarized in Table 1. The parity products for the occupied states at the TRIM points are calculated by $\delta(k_i) = \prod_{N=1}^{12} \xi(i)$, to be $+1$, $-1$, $+1$, and $+1$ for $\Gamma$, X, S, and Y, respectively. This yields a nontrivial topological invariant $\nu = 1$ by $(-1)^\nu = \prod_{i=1}^4 \delta(k_i)$. This result indicates the presence of nontrivial topological Dirac semimetal in $C_2(BN)_2$ with $\gamma = 95.5°$.

TABLE 1. The calculated parity eigenvalues of the 12 occupied bands at four TRIM points for $C_2(BN)_2$ biphenylene with $\gamma = 95.5°$.

| TRIM | Parities | Product |
|---|---|---|
| $\Gamma$ | + − + + − − + + − + + − | + |
| X | − + − + + − − + + − + + | − |
| S | − + − + − + − + − + − + | + |
| Y | + − + − + − − − + + − + | + |

## B. Virtual-crystal approximation and tight-binding model

To understand why the DPs in the $C_2(BN)_2$ system are located in non-HSR after substituting part of the carbon-carbon bonds in biphenylene with the boron-nitrogen bonds. We studied the energy structures [Fig. 6(a-d)]and the positions [Fig. 6(e)] of the DPs as a function of composition $t$ for the whole range ($t$=0 to 1) based on the first-principles virtual-crystal approximation (VCA) [60-61] calculations for the $C_{6-4t}(BN)_{2t}$ BPN structure. We found that the DPs are located on the high symmetric line $\Gamma$-Y with the $t$=0, and with the doping of B and N, the DPs deviate from the high symmetric line and move towards the non-high symmetric region. These DPs pass through the high symmetry line Γ-S [Fig. 6(i)] at the t=0.90. When the boron and nitrogen doping ratio is 82%, the conduction band and valence band cross at the TRIM point X to produce the type-I Dirac point, and then this DP moves away from the point X to the non-HSR accompanied by the band inversion as shown in Fig. 6(g-j). As the doping ratio continues to increase, the two pairs of Dirac points approach each other and eventually annihilate in non-HSR. The annihilation of two pairs of Dirac points in non-HSR causes two corresponding edge states stick together to generate a stripe of isolated edge state between the conduction and the valence band, which is consistent with the inverse process of the evolution of the edge states [Fig. 4(a-f)] under the regulation of strains for $C_2(BN)_2$. Fig. 6(f) shows a schematic diagram of the movement of the Dirac points in the Brillouin region during this process of substituting.

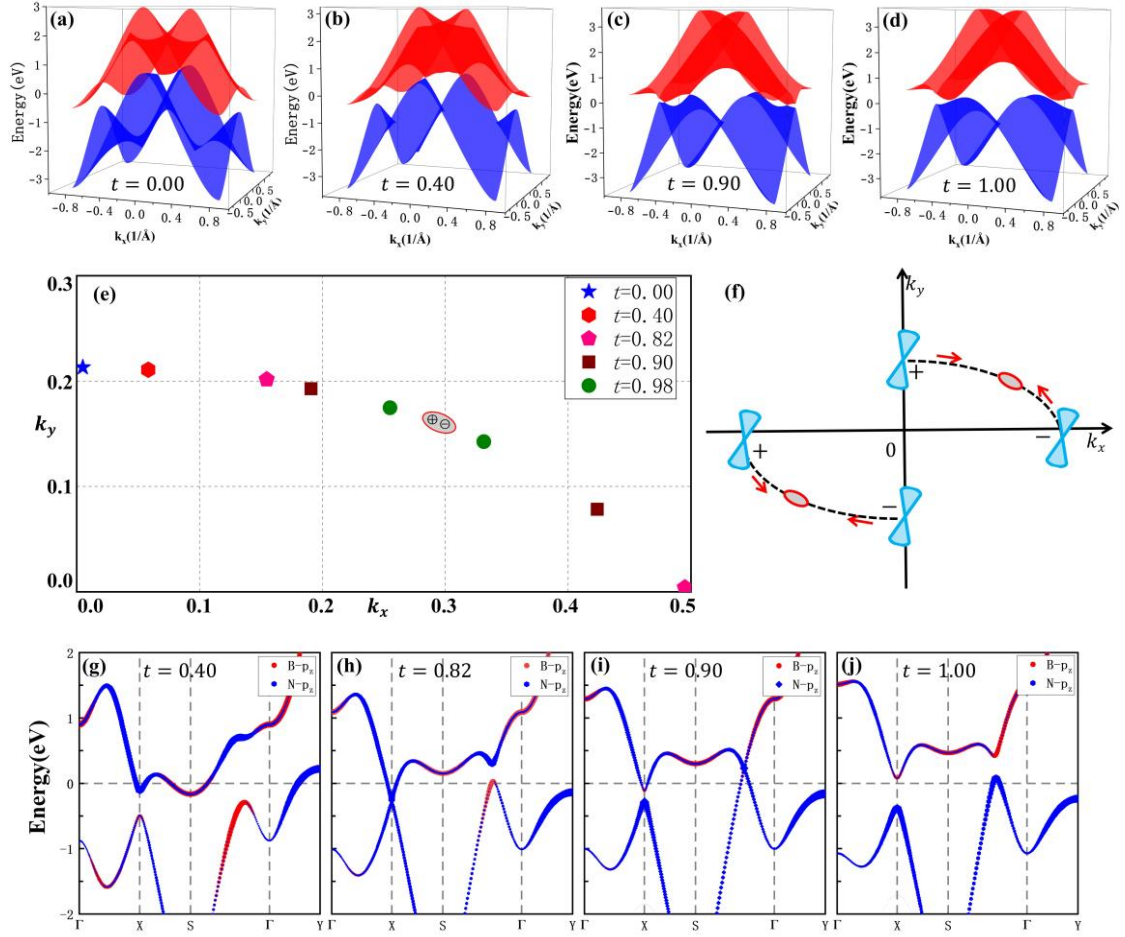

**Fig 6.** (a-d) The electronic band structure for the conduction band and valence band in the 2D Brillouin zone with different doping ratio $t$. (e) The shift of the location of the Dirac points in the first quadrant of Brillouin zone. (f) The schematic diagram of the movement of the Dirac points in the Brillouin region during this process of substituting. (g-h) the orbital-resolved bulk band structure with different doping ratio $t$.

Hereafter, we discussed the mechanism of the deviation of the DPs and the evolution of the corresponding edge states from the system of C-BPN to $C_2(BN)_2$ by using a simple TB model. All topological features found in our *ab initio* virtual-crystal calculations can be retrieved by TB approximations with the $p_z$ orbitals. Given the presence of three distinct types of atoms for the system of $C_2(BN)_2$, we set different hopping energies ($t_\alpha$, $t_\beta$, and $t_\gamma$) for each of the n.n. pair. Meanwhile, we considered the site energies ($\epsilon_B$, $\epsilon_C$, and $\epsilon_N$) for the different atoms as well as the hopping of the next n.n. pairs in the square ($t_\delta$ and $t_\varepsilon$) and hexagonal ($t_\epsilon$ and $t_\zeta$) rings to obtain the energy band structure which is similar to the DFT bands. With these parameters, a TB model can be written as

$$\mathcal{H}_{TB} = -t_0 \sum_{\langle i,j \rangle} c_i^\dagger c_j - t_d \sum_{\langle\langle i,j \rangle\rangle \in S} c_i^\dagger c_j - t_p \sum_{\langle\langle i,j \rangle\rangle \in H} c_i^\dagger c_j + \epsilon \sum c_i^\dagger c_i + (c.c.) \quad (7)$$

$$t_0 = \begin{cases} t_\alpha, for\ \langle i,j \rangle \in \{C-C; B-N\} \\ t_\beta, for\ \langle i,j \rangle \in \{C-B\} \\ t_\gamma, for\ \langle i,j \rangle \in \{C-N\} \end{cases};$$

$$t_d = \begin{cases} t_\delta, for\ \langle\langle i,j \rangle\rangle \in \{B-B\} \\ t_\varepsilon, for\ \langle\langle i,j \rangle\rangle \in \{N-N\} \end{cases};$$

$$t_p = \begin{cases} t_\epsilon, for\ \langle\langle i,j \rangle\rangle \in \{C-B; C-N\} \\ t_\zeta, for\ \langle\langle i,j \rangle\rangle \in \{B-N\} \end{cases};$$

where $S$ and $H$ denote the atoms belonging to the square and hexagonal rings, respectively. The $\langle \cdots \rangle$ and $\langle\langle \cdots \rangle\rangle$ indicate the n.n. and next n.n. pairs, respectively. The scheme of the hopping amplitudes for the $C_2(BN)_2$ system is shown in the Fig. S3 in SM. For the system of C-BPN, without any loss of generality, we set the same hopping energies with the Ref. [20] to reproduce the topological properties [Fig. 7(a)]. The edge state only exists in the armchair ribbons [19-20] with width N=50 (see Fig. S4 in SM).

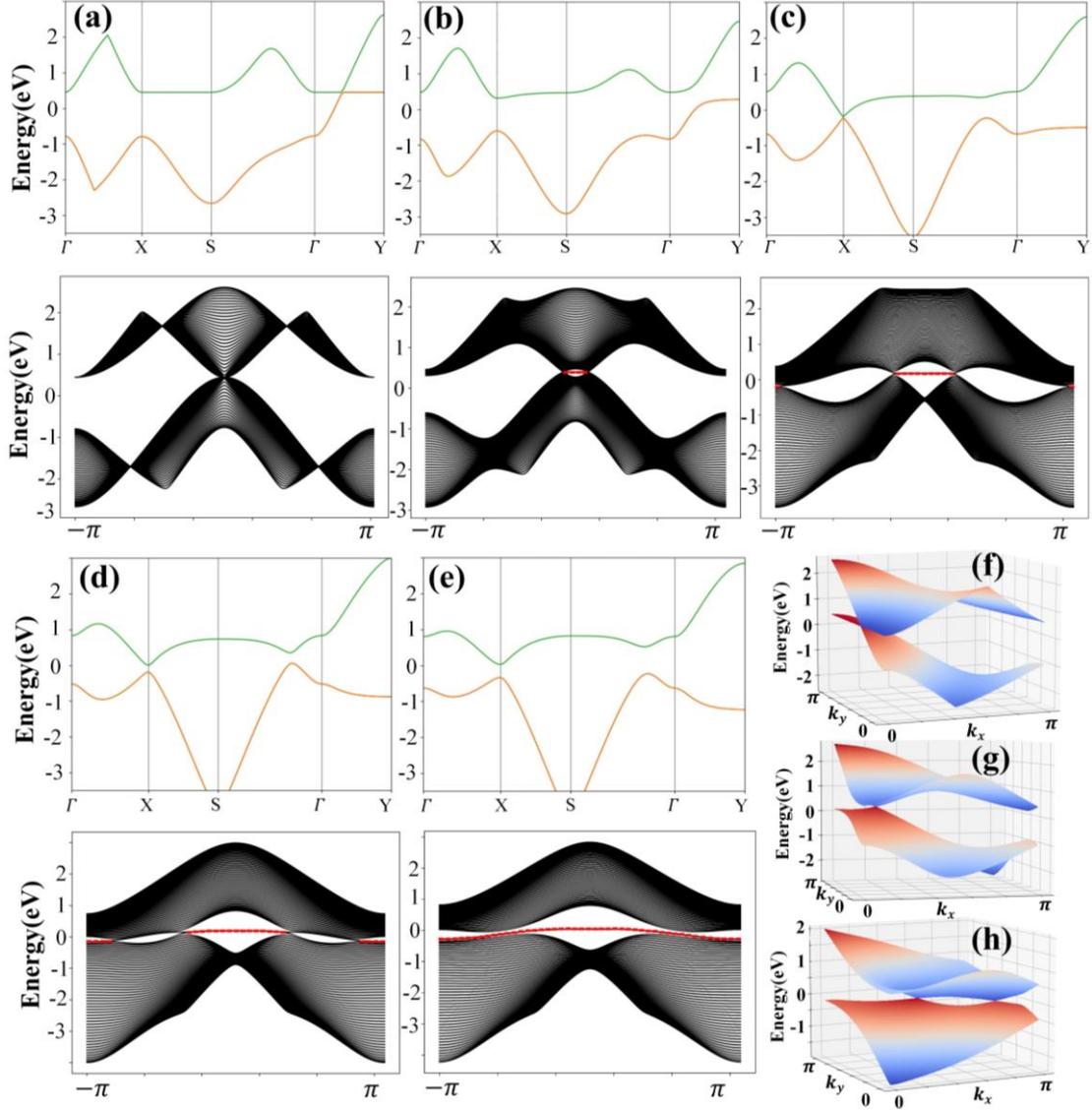

**Fig 7.** (a-e) The band structure of bulk and zigzag ribbons with width N=50. The red dashed lines represent the edge states. (f-h) The 3D band structure corresponding to Fig 7(a), (b), (d), respectively. The setting of parameters is summarized in Table S2 in the SM.

For the system of the $C_2(BN)_2$ network, the band structures of bulk and zigzag ribbons are shown in Fig. 7(e), which were calculated by the TB model of eq 7 with different parameters that were set according to the TB Hamiltonian constructed by using the *WANNIER90* package. Significantly, the TB model of this system with inversion symmetry has broken the mirror symmetry in comparison to the system of C-BPN. As shown in Fig. 7(e) and Fig. 6(j), although the energy band structure is not same with the DFT band at the point of $\Gamma$ and S, the difference has no effect on the topological properties [Fig. 2(d)]. To study the mechanism of the deviation of the DPs from the HSR and the evolution of the corresponding edge states, we performed interpolation between the parameters of pure carbon and BCN systems. As shown in Fig. 7(f, g), we found that the changing of hopping energies $t_d$ (from $t_\delta = t_\varepsilon = 0.45$ eV to $t_\delta = 0.95$ eV and $t_\varepsilon = -0.10$ eV) for the next n.n. pairs in the square ring is the main reason that the DPs shift from the high-symmetry line $\Gamma$-Y to the non-HSR and the corresponding edge state[Fig. 7(b)] emerges on the zigzag ribbons. With the change of parameters, a new pair of DPs [Fig. 7(c)] appears at the TRIM point X and then moves to the non-HSR [Fig. 7(h)]. The emergence and evolution of the corresponding edge states can be observed in Fig. 7(c-d). In the end, the two pairs of DPs approach each other and eventually annihilate in the non-HSR. At the same time, the two corresponding edge states stick together to form an isolated edge state [Fig. 7(d-e)], which is consistent with the DFT calculations. The topological characteristics and edge states could also be verified by calculating the Zak phase. Here, we have calculated the Zak phase (Fig. 8) of the upper valence band along the $k_y$ direction for each $k_x$,

$\gamma_v(k_x) = i \int_{-\pi/b}^{\pi/b} dk_y \left\langle u_v(\boldsymbol{k}) \middle| \partial_{k_y} \middle| u_v(\boldsymbol{k}) \right\rangle$, where the $|u_v(\boldsymbol{k})\rangle$ represents the cell periodic function of the upper valence band Bloch state of eq 7. The $\gamma_v(k_x)$ is defined modulo $2\pi$ and should be quantized into 0 or $\pi$ due to the chiral symmetry[62-63] or

inversion symmetry[64]. The value of the Zak phase $\gamma_v(k_x)$ is equal to $\pi$, which means that one can guarantee the edge states exist in this region [65]. Consistently, the Zak phase changes from 0 (line-a in Fig. 8) for the C-BPN with chiral symmetry to local $\pi$ at $\Gamma$ and Y points [line-(b-d) in Fig. 8] for the $C_2(BN)_2$ with inversion symmetry as the DPs shift away from the high-symmetry line and emerge at the X point. Finally, the two pairs of DPs are close to each other and annihilate to the non-HSR, resulting a Zak phase of $\pi$ (line-e in Fig. 8).

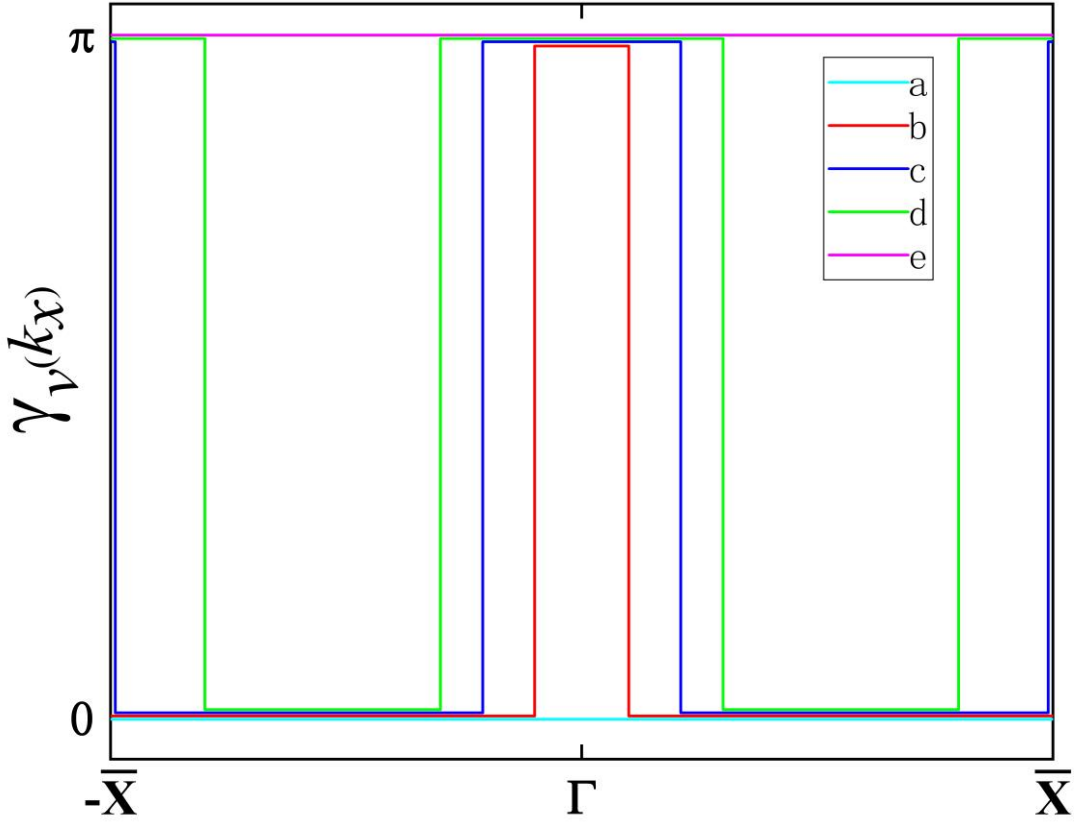

**Fig 8.** The calculated Zak phase $\gamma_v(k_x)$ of the upper valence band along the $k_x$ direction. The lines (a-e) represent the same set of parameters in Fig. 7(a-e), respectively.

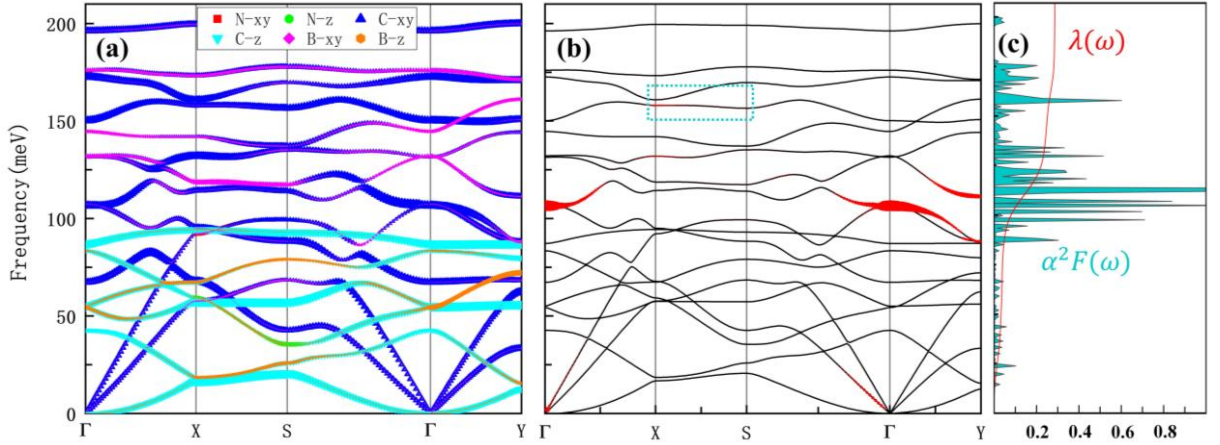

**Fig 9.** (a) Phonon dispersions weighted by the vibration modes, (b) phonon dispersion weighted by magnitude of EPC $\lambda_{qv}$ for $C_4BN$ biphenylene, dashed box indicates the main $\lambda_{qv}$ difference between the $C_4BN$ and C-BPN. (c) Frequency dependence of the Eliashberg spectral function $\alpha^2F(\omega)$ and the cumulative EPC $\lambda(\omega)$ for $C_4BN$.

## C. Superconducting properties of $C_4BN$

To investigate the superconductivity of the $C_{6-2x}(BN)_x$ biphenylene networks, we have calculated the phonon dispersions weighted by the vibration mode and by the magnitude of EPC $\lambda_{qv}$, respectively, as shown in Fig. 9(a, b). Figure 9(c) shows the Eliashberg spectral function $\alpha^2F(\omega)$ and the EPC constant $\lambda(\omega)$ over the whole frequency range. Since none of the $C_{6-2x}(BN)_x$ biphenylene exhibits superconducting, except $C_4BN$, we only show the results of $C_4BN$. As shown in Fig. 9(a, b), the dominant contribution to EPC [red circles in Fig. 9(b)] originates from the in-plane vibrational mode of C atoms of the square ring. The majority of the strong peaks of the Eliashberg spectral function $\alpha^2F(\omega)$ also concentrate in the mid-frequency zone [see Fig. 9(c)]. From the Eliashberg spectral function, there is a strong EPC as indicated by the peak around 113meV, which is consistent with Fig. 9(b). From the accumulative EPC constant $\lambda(\omega)$ shown in Fig. 9(c), the $\lambda$ from the mid-frequency modes between 85meV to 125meV is 0.18, contributing 63% of the total EPC constant ($\lambda = 0.286$). The value of $\lambda$ is slightly smaller than the C-BPN[19], resulting a lower superconducting temperature Tc=0.38K, compared to Tc=0.59K of C-BPN[19] with the same $\mu^* = 0.1$. The possible reason is that the EPC strength in C$_4$BN is weakened at higher phonon frequency about 160meV [see the dashed box in Fig. 9(b)], due to the doping of boron and nitrogen. To further justify the above argument, we have calculated the mode integrated EPC $\lambda_{tot}(\boldsymbol{q}) = \sum_\nu \lambda_\nu(\boldsymbol{q})$ along high symmetry lines (see Fig. 10). The average value of the EPC strength along X-S is about 0.1, which is significantly lower than 0.2 for the C-BPN[19].

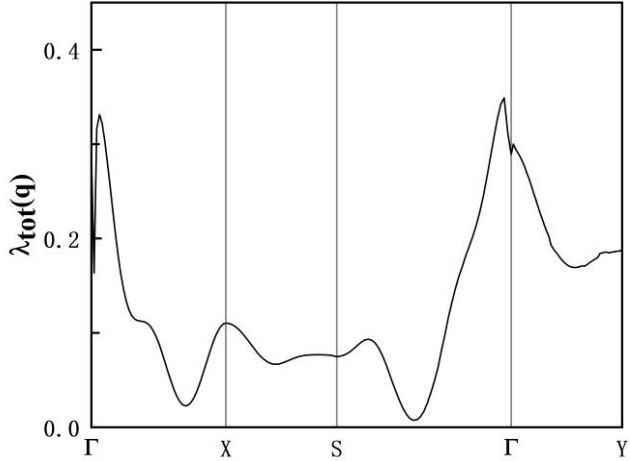

**Fig 10.** The EPC $\lambda_{tot}(\boldsymbol{q})$ along high symmetry directions for C$_4$BN.

## IV. Conclusions

In summary, we constructed two different stable configurations of the C$_{6-2x}$(BN)$_x$ biphenylene networks and investigated their corresponding electronic structure and edge states in great details. In C$_4$BN, we discovered the paired type-II Dirac cones, which is similar to the C-BPN. In $C_2(BN)_2$, we observed an isolated complete edge state between the conduction and valence bands. By regulation of the cell angel $\gamma$, we found two pairs of Dirac points appear on the non-HSR and the corresponding complete edge state is split into two parts of edge states. With the $\gamma = 95.5°$, a pair of Dirac points with opposite chirality annihilates at the point X of TRIM, which produces a topological Dirac semimetal phase with a nontrivial $\mathbb{Z}_2$ topological invariant. By the VCA and TB model analysis, we found hopping along the diagonal direction in the square ring might be the main reason why the DPs shift away from the HSR to the non-HSR. We illustrated that due to the replacement of carbon atoms by boron and nitrogen, which leads to a weakened EPC strength in $C_4BN$ compared to C-BPN. This study could propose a new understanding of the topological properties in the 2D Dirac semimetals.

## Acknowledgements

This work was supported by National Key R&D Program of China under Grant No. 2021YFB3802300, the National Natural Science Foundation of China under Grant No. 11672274 and the NSAF under Grant No. U1730248. Part of the computation was performed using the supercomputer at the Center for Computational Materials Science (CCMS) of the Institute for Materials Research (IMR) at Tohoku University, Japan.

## References

[1] Novoselov K S, Geim A K, Morozov S V, et al. Electric field effect in atomically thin carbon films[J]. science, 2004, 306(5696): 666-669. https://www.science.org/doi/10.1126/science.1102896

[2] Geim A K. Graphene: status and prospects[J]. science, 2009, 324(5934): 1530-1534.

https://www.science.org/doi/10.1126/science.1158877.

[3] Neto A H C, Guinea F, Peres N M R, et al. The electronic properties of graphene[J]. Reviews of modern physics, 2009, 81(1): 109.

https://www.science.org/doi/10.1126/science.1158877

[4] Avouris P. Graphene: electronic and photonic properties and devices[J]. Nano letters, 2010, 10(11): 4285-4294. https://doi.org/10.1021/nl102824h.

[5] Sarma S D, Adam S, Hwang E H, et al. Electronic transport in two-dimensional graphene[J]. Reviews of modern physics, 2011, 83(2): 407.

https://doi.org/10.1103/RevModPhys.83.407.

[6] Balandin A A, Ghosh S, Bao W, et al. Superior thermal conductivity of single-layer graphene[J]. Nano letters, 2008, 8(3): 902-907. https://doi.org/10.1021/nl0731872.

[7] Balandin A A. Thermal properties of graphene and nanostructured carbon materials[J]. Nature materials, 2011, 10(8): 569-581.

https://www.nature.com/articles/nmat3064.

[8] Pop E, Varshney V, Roy A K. Thermal properties of graphene: Fundamentals and applications[J]. MRS bulletin, 2012, 37(12): 1273-1281.

https://doi.org/10.1557/mrs.2012.203.

[9] Papageorgiou D G, Kinloch I A, Young R J. Mechanical properties of graphene and graphene-based nanocomposites[J]. Progress in materials science, 2017, 90: 75-127.

https://doi.org/10.1016/j.pmatsci.2017.07.004.

[10] Frank I W, Tanenbaum D M, van der Zande A M, et al. Mechanical properties of suspended graphene sheets[J]. Journal of Vacuum Science & Technology B:

Microelectronics and Nanometer Structures Processing, Measurement, and Phenomena, 2007, 25(6): 2558-2561. https://doi.org/10.1116/1.2789446.

[11] Haldane F D M. Model for a quantum Hall effect without Landau levels: Condensed-matter realization of the" parity anomaly"[J]. Physical review letters, 1988, 61(18): 2015. https://doi.org/10.1103/PhysRevLett.61.2015.

[12] Kane C L, Mele E J. $Z_2$ topological order and the quantum spin Hall effect[J]. Physical review letters, 2005, 95(14): 146802.

https://doi.org/10.1103/PhysRevLett.95.146802.

[13] Thouless D J, Kohmoto M, Nightingale M P, et al. Quantized Hall conductance in a two-dimensional periodic potential[J]. Physical review letters, 1982, 49(6): 405.

https://doi.org/10.1103/PhysRevLett.49.405.

[14] Bernevig B A, Zhang S C. Quantum spin Hall effect[J]. Physical review letters, 2006, 96(10): 106802. https://doi.org/10.1103/PhysRevLett.96.106802.

[15] Cortijo A, Guinea F, Vozmediano M A H. Geometrical and topological aspects of graphene and related materials[J]. Journal of Physics A: Mathematical and Theoretical, 2012, 45(38): 383001. https://doi.org/10.1088/1751-8113/45/38/383001.

[16] Gmitra M, Konschuh S, Ertler C, et al. Band-structure topologies of graphene: Spin-orbit coupling effects from first principles[J]. Physical Review B, 2009, 80(23): 235431. https://doi.org/10.1103/PhysRevB.80.235431.

[17] Fan Q, Yan L, Tripp M W, et al. Biphenylene network: A nonbenzenoid carbon allotrope[J]. Science, 2021, 372(6544): 852-856.

https://www.science.org/doi/10.1126/science.abg4509.

[18] Tyutyulkov N, Dietz F, Müllen K, et al. Structure and energy spectra of a two-dimensional dielectric carbon allotrope[J]. Chemical physics letters, 1997, 272(1-2): 111-114. https://doi.org/10.1016/S0009-2614(97)00465-X.

[19] Liu P F, Li J, Zhang C, et al. Type-II Dirac cones and electron-phonon interaction in monolayer biphenylene from first-principles calculations[J]. Physical Review B, 2021, 104(23): 235422. https://doi.org/10.1103/PhysRevB.104.235422.

[20] Son Y W, Jin H, Kim S. Magnetic Ordering, Anomalous Lifshitz Transition, and Topological Grain Boundaries in Two-Dimensional Biphenylene Network[J]. Nano Letters, 2022, 22(7): 3112-3117. https://doi.org/10.1021/acs.nanolett.2c00528.

[21] Yang N, Yang H, Jin G. Interface-induced topological phase and doping-modulated bandgap of two-dimensioanl graphene-like networks[J]. Chinese Physics B, 2023, 32(1): 017201. https://doi.org/10.1088/1674-1056/ac904d.

[22] Farzadian O, Dehaghani M Z, Kostas K V, et al. A theoretical insight into phonon heat transport in graphene/biphenylene superlattice nanoribbons: A molecular dynamic study[J]. Nanotechnology, 2022, 33(35): 355705.

https://doi.org/10.1088/1361-6528/ac733e.

[23] Mashhadzadeh A H, Dehaghani M Z, Molaie F, et al. A theoretical insight into the mechanical properties and phonon thermal conductivity of biphenylene network structure[J]. Computational Materials Science, 2022, 214: 111761.

https://doi.org/10.1016/j.commatsci.2022.111761.

[24] Tong Z, Pecchia A, Yam C Y, et al. Ultrahigh Electron Thermal Conductivity in T-Graphene, Biphenylene, and Net-Graphene[J]. Advanced Energy Materials, 2022,

12(28): 2200657. https://doi.org/10.1002/aenm.202200657.

[25] Xie Z X, Chen X K, Yu X, et al. Intrinsic thermoelectric properties in biphenylene nanoribbons and effect of lattice defects[J]. Computational Materials Science, 2023, 220: 112041. https://doi.org/10.1016/j.commatsci.2023.112041.

[26] Sahoo M R, Ray A, Ahuja R, et al. Activation of metal-free porous basal plane of biphenylene through defects engineering for hydrogen evolution reaction[J]. International Journal of Hydrogen Energy, 2023, 48(28): 10545-10554.

https://doi.org/10.1016/j.ijhydene.2022.11.298.

[27] Ge Y, Wang Z, Wang X, et al. Superconductivity in the two-dimensional nonbenzenoid biphenylene sheet with Dirac cone[J]. 2D Materials, 2021, 9(1): 015035.

https://doi.org/10.1088/2053-1583/ac4573.

[28] Chakraborty H, Mogurampelly S, Yadav V K, et al. Phonons and thermal conducting properties of borocarbonitride (BCN) nanosheets[J]. Nanoscale, 2018, 10(47): 22148-22154. https://doi.org/10.1039/C8NR07373B.

[29] Demirci S, Çallıoğlu Ş, Görkan T, et al. Stability and electronic properties of monolayer and multilayer structures of group-IV elements and compounds of complementary groups in biphenylene network[J]. Physical Review B, 2022, 105(3): 035408. https://doi.org/10.1103/PhysRevB.105.035408.

[30] Feng Z, Zhang B, Li R, et al. Biphenylene with doping B/N as promising metal-free single-atom catalysts for electrochemical oxygen reduction reaction[J]. Journal of Power Sources, 2023, 558: 232613. https://doi.org/10.1016/j.jpowsour.2022.232613.

[31] Gorkan T, Çallıoğlu S, Demirci S, et al. Functional Carbon and Silicon Monolayers

in Biphenylene Network[J]. ACS Applied Electronic Materials, 2022, 4(6): 3056-3070.

https://doi.org/10.1021/acsaelm.2c00459.

[32] Liu G, Guo A, Cao F, et al. Lattice Thermal Conductivity of Silicon Monolayer in Biphenylene Network[J]. Available at SSRN 4355188.

https://doi.org/10.1063/5.0155409.

[33] Liu G H, Yang L, Qiao S X, et al. Superconductivity of monolayer functionalized biphenylene with Dirac cones[J]. Physical Chemistry Chemical Physics, 2023.

https://doi.org/10.1039/D2CP04381E.

[34] Lv F, Liang H, Duan Y. Funnel-shaped electronic structure and enhanced thermoelectric performance in ultralight $C_x(BN)_{1-x}$ biphenylene networks[J]. Physical Review B, 2023, 107(4): 045422. https://doi.org/10.1103/PhysRevB.107.045422.

[35] Ma X D, Tian Z W, Jia R, et al. BN counterpart of biphenylene network: A theoretical investigation[J]. Applied Surface Science, 2022, 598: 153674.

https://doi.org/10.1016/j.apsusc.2022.153674.

[36] Ren K, Shu H, Huo W, et al. Tuning electronic, magnetic and catalytic behaviors of biphenylene network by atomic doping[J]. Nanotechnology, 2022, 33(34): 345701.

https://doi.org/10.1088/1361-6528/ac6f64.

[37] Dehaghani M Z, Farzadian O, Kostas K V, et al. Theoretical study of heat transfer across biphenylene/h-BN superlattice nanoribbons[J]. Physica E: Low-dimensional Systems and Nanostructures, 2022, 144: 115411.

https://doi.org/10.1016/j.physe.2022.115411.

[38] Zhang X, Chen F, Jia B, et al. A novel lithium decorated N-doped 4, 6, 8-

biphenylene for reversible hydrogen storage: Insights from density functional theory[J]. International Journal of Hydrogen Energy, 2023, 48(45): 17216-17229. https://doi.org/10.1016/j.ijhydene.2023.01.222.

[39] Bafekry A, Naseri M, Fadlallah M M, et al. A novel two-dimensional boron–carbon–nitride (BCN) monolayer: A first-principles insight[J]. Journal of Applied Physics, 2021, 130(11): 114301. https://doi.org/10.1063/5.0062323.

[40] Mortazavi B. Ultrahigh thermal conductivity and strength in direct-gap semiconducting graphene-like $BC_6N$: A first-principles and classical investigation[J]. Carbon, 2021, 182: 373-383. https://doi.org/10.1016/j.carbon.2021.06.038.

[41] Mortazavi B, Shojaei F, Yagmurcukardes M, et al. Anisotropic and outstanding mechanical, thermal conduction, optical, and piezoelectric responses in a novel semiconducting BCN monolayer confirmed by first-principles and machine learning[J]. Carbon, 2022, 200: 500-509. https://doi.org/10.1016/j.carbon.2022.08.077.

[42] Nehate S D, Saikumar A K, Prakash A, et al. A review of boron carbon nitride thin films and progress in nanomaterials[J]. Materials Today Advances, 2020, 8: 100106. https://doi.org/10.1016/j.mtadv.2020.100106.

[43] Tang Z, Cruz G J, Wu Y, et al. Giant Narrow-Band Optical Absorption and Distinctive Excitonic Structures of Monolayer $C_3N$ and $C_3B$[J]. Physical Review Applied, 2022, 17(3): 034068. https://doi.org/10.1103/PhysRevApplied.17.034068.

[44] Wu Y, Chen Y, Ma C, et al. Monolayer $C_7N_6$: Room-temperature excitons with large binding energies and high thermal conductivities[J]. Physical Review Materials, 2020, 4(6): 064001. https://doi.org/10.1103/PhysRevMaterials.4.064001.

[45] Kohn W, Sham L J. Self-consistent equations including exchange and correlation effects[J]. Physical review, 1965, 140(4A): A1133.

https://doi.org/10.1103/PhysRev.140.A1133.

[46] Hohenberg P, Kohn W. Inhomogeneous electron gas[J]. Physical review, 1964, 136(3B): B864. https://doi.org/10.1103/PhysRev.136.B864.

[47] Kresse G, Furthmüller J. Efficient iterative schemes for ab initio total-energy calculations using a plane-wave basis set[J]. Physical review B, 1996, 54(16): 11169.

https://doi.org/10.1103/PhysRevB.54.11169.

[48] Perdew J P, Burke K, Ernzerhof M. Generalized gradient approximation made simple[J]. Physical review letters, 1996, 77(18): 3865.

https://doi.org/10.1103/PhysRevLett.77.3865.

[49] Kresse G, Joubert D. From ultrasoft pseudopotentials to the projector augmented-wave method[J]. Physical review b, 1999, 59(3): 1758.

https://doi.org/10.1103/PhysRevB.59.1758.

[50] Monkhorst H J, Pack J D. Special points for Brillouin-zone integrations[J]. Physical review B, 1976, 13(12): 5188. https://doi.org/10.1103/PhysRevB.13.5188.

[51] Marzari N, Mostofi A A, Yates J R, et al. Maximally localized Wannier functions: Theory and applications[J]. Reviews of Modern Physics, 2012, 84(4): 1419.

https://doi.org/10.1103/RevModPhys.84.1419.

[52] Mostofi A A, Yates J R, Pizzi G, et al. An updated version of wannier90: A tool for obtaining maximally-localised Wannier functions[J]. Computer Physics Communications, 2014, 185(8): 2309-2310. https://doi.org/10.1016/j.cpc.2014.05.003.

[53] Pizzi G, Vitale V, Arita R, et al. Wannier90 as a community code: new features and applications[J]. Journal of Physics: Condensed Matter, 2020, 32(16): 165902. https://doi.org/10.1088/1361-648X/ab51ff.

[54] Wu Q S, Zhang S N, Song H F, et al. WannierTools: An open-source software package for novel topological materials[J]. Computer Physics Communications, 2018, 224: 405-416. https://doi.org/10.1016/j.cpc.2017.09.033.

[55] Troullier N, Martins J L. Efficient pseudopotentials for plane-wave calculations[J]. Physical review B, 1991, 43(3): 1993. https://doi.org/10.1103/PhysRevB.43.1993.

[56] Giannozzi P, Baroni S, Bonini N, et al. QUANTUM ESPRESSO: a modular and open-source software project for quantum simulations of materials[J]. Journal of physics: Condensed matter, 2009, 21(39): 395502.

[57] Giustino F. Electron-phonon interactions from first principles[J]. Reviews of Modern Physics, 2017, 89(1): 015003. https://doi.org/10.1103/RevModPhys.89.015003.

[58] Allen P B, Dynes R C. Transition temperature of strong-coupled superconductors reanalyzed[J]. Physical Review B, 1975, 12(3): 905. https://doi.org/10.1103/PhysRevB.12.905.

[59] Eliashberg G M. Interactions between electrons and lattice vibrations in a superconductor[J]. Sov. Phys. JETP, 1960, 11(3): 696-702.

[60] Bellaiche L, Vanderbilt D. Virtual crystal approximation revisited: Application to dielectric and piezoelectric properties of perovskites[J]. Physical Review B, 2000, 61(12): 7877. https://doi.org/10.1103/PhysRevB.61.7877.

[61] Eckhardt C, Hummer K, Kresse G. Indirect-to-direct gap transition in strained and unstrained Sn x Ge 1− x alloys[J]. Physical Review B, 2014, 89(16): 165201. https://doi.org/10.1103/PhysRevB.89.165201.

[62] Hatsugai Y. Bulk-edge correspondence in graphene with/without magnetic field: Chiral symmetry, Dirac fermions and edge states[J]. Solid state communications, 2009, 149(27-28): 1061-1067. https://doi.org/10.1016/j.ssc.2009.02.055.

[63] Lau A, Van Den Brink J, Ortix C. Topological mirror insulators in one dimension[J]. Physical Review B, 2016, 94(16): 165164. https://doi.org/10.1103/PhysRevB.94.165164.

[64] van Miert G, Ortix C, Smith C M. Topological origin of edge states in two-dimensional inversion-symmetric insulators and semimetals[J]. 2D Materials, 2016, 4(1): 015023. https://doi.org/10.1088/2053-1583/4/1/015023.

[65] Ryu S, Hatsugai Y. Topological origin of zero-energy edge states in particle-hole symmetric systems[J]. Physical review letters, 2002, 89(7): 077002. https://doi.org/10.1103/PhysRevLett.89.077002.

# Supplementary Materials for "Topological and superconducting properties of two-dimensional $C_{6-2x}(BN)_x$ biphenylene network: a first-principles investigation"

Guang F. Yang,[1] Hong X. Song,[1,*] Dan Wang,[1] Hao Wang,[1] and Hua Y. Geng[1,2,*]

[1]*National Key Laboratory of Shock Wave and Detonation Physics, Institute of Fluid Physics, China Academy of Engineering Physics, Mianyang, Sichuan 621900, P. R. China;*

[2]*HEDPS, Center for Applied Physics and Technology, and College of Engineering, Peking University, Beijing 100871, P. R. China;*

* Corresponding authors:
E-mail address: s102genghy@caep.cn (Hua Y. Geng), hxsong555@163.com (Hong X. Song).

**Supporting Figures and Tables**

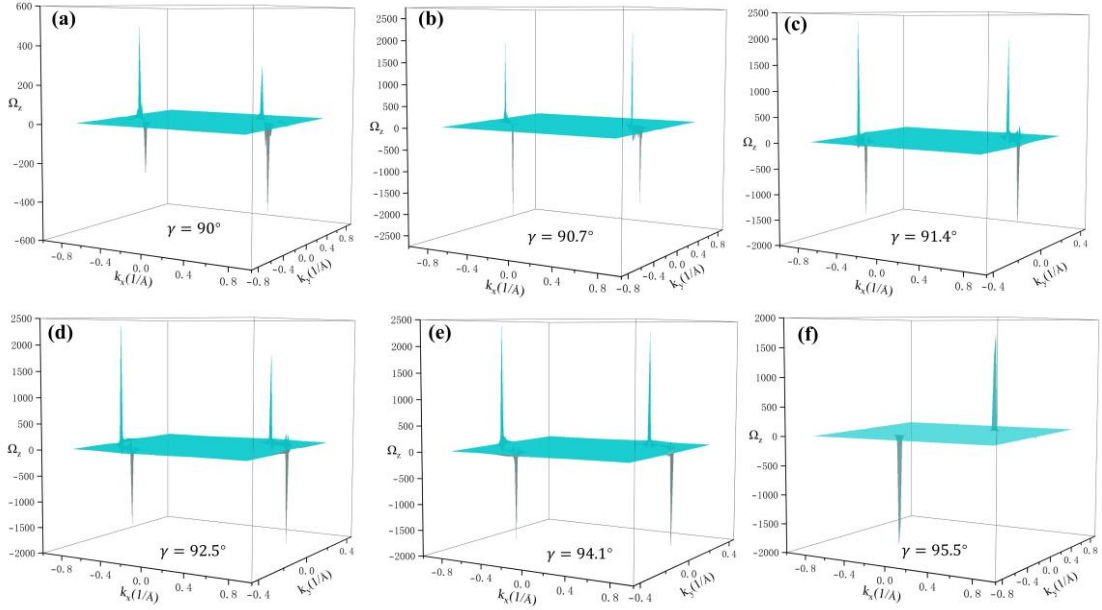

**FIG. S1.** The Berry curvature $\Omega_z$ of the $C_2(BN)_2$ with different cell angle $\gamma$.

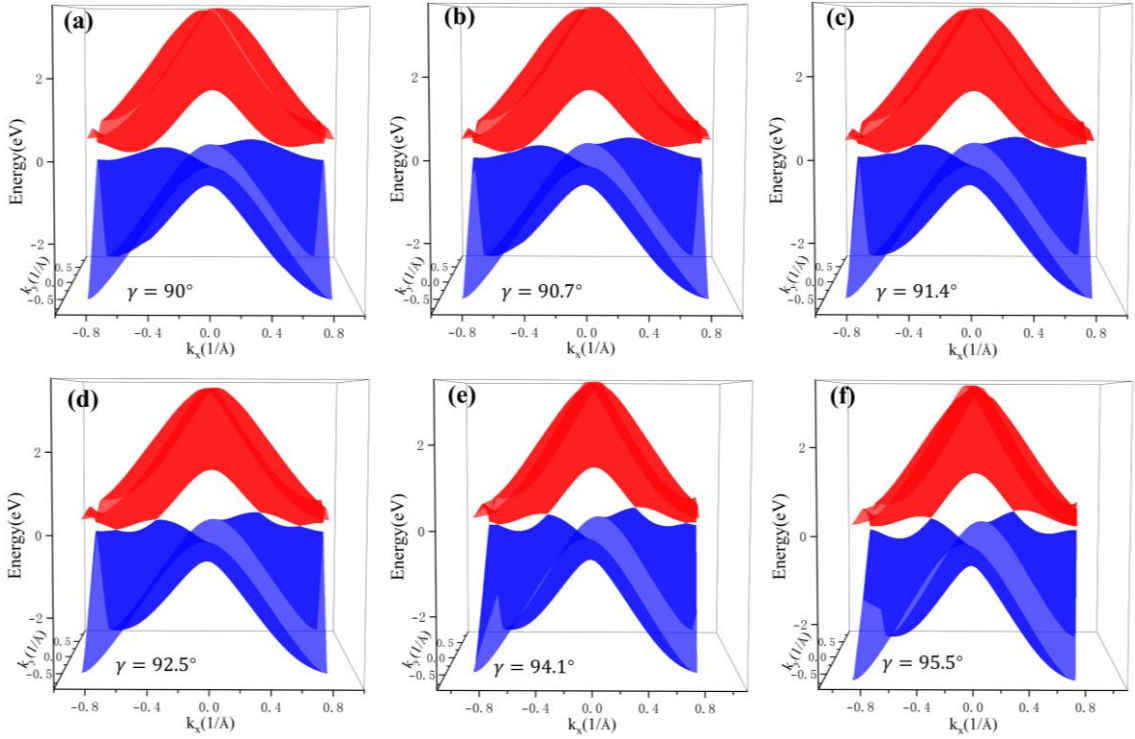

**FIG. S2.** The 3D electronic band structure of the $C_2(BN)_2$ with different cell angle $\gamma$.

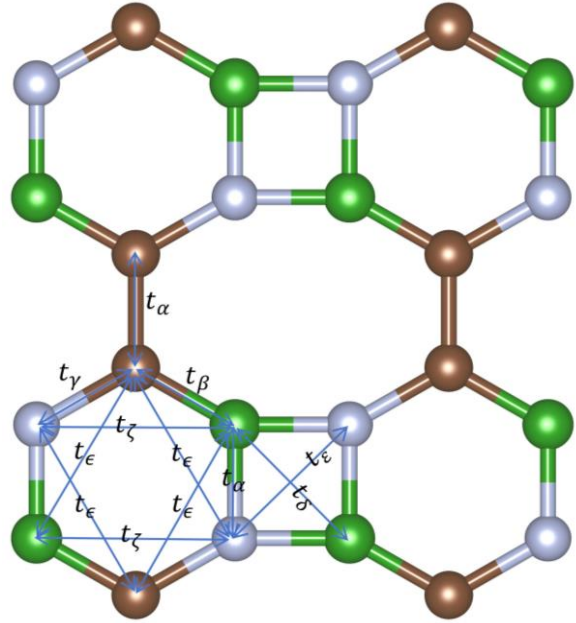

**FIG. S3.** The scheme of the hopping amplitudes for the tight-binding model of eq 7 in

the $C_2(BN)_2$ network

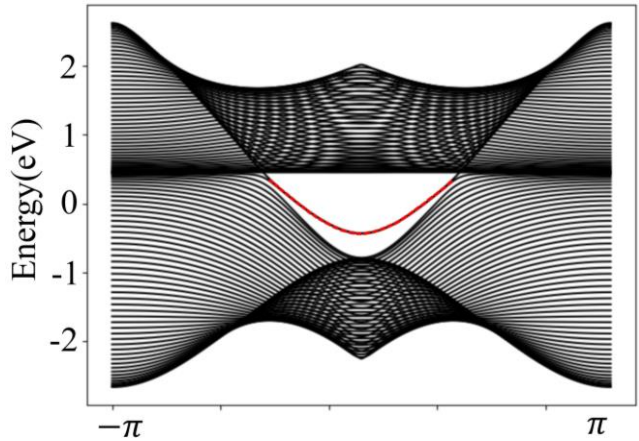

**FIG. S4.** The band structure of the armchair ribbons with width N=50.

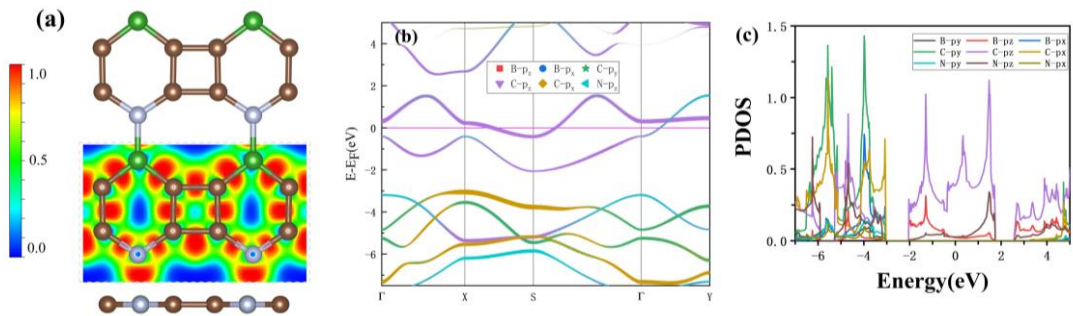

**FIG. S5.** The atomic structure and electron localization function(a), orbital-resolved

bulk band structure(b) and projected DOS(c) of $C_4BN$.

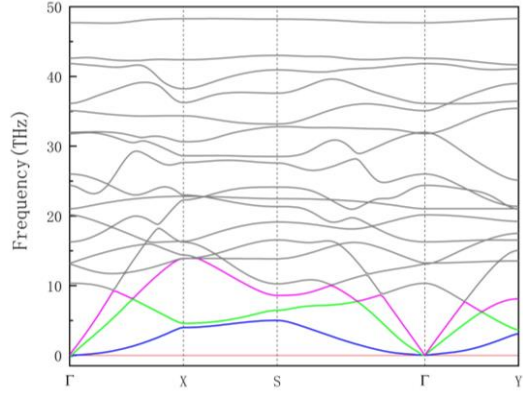

**FIG. S6.** Phonon band spectrum of the structure of C$_4$BN.

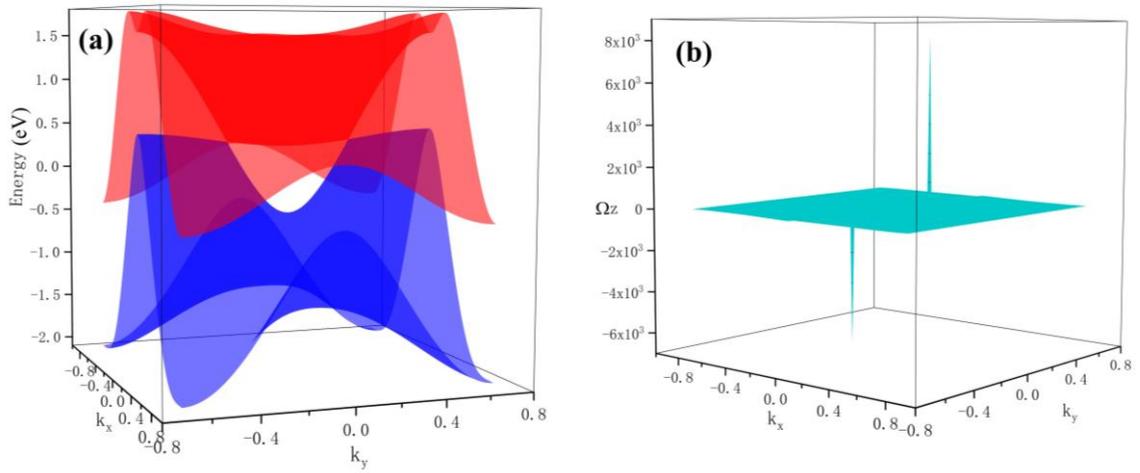

**FIG. S7.** Electronic band structure (a) and the Berry curvature $\Omega_z$ (b) in the 2D

Brillouin zone of the C$_4$BN.

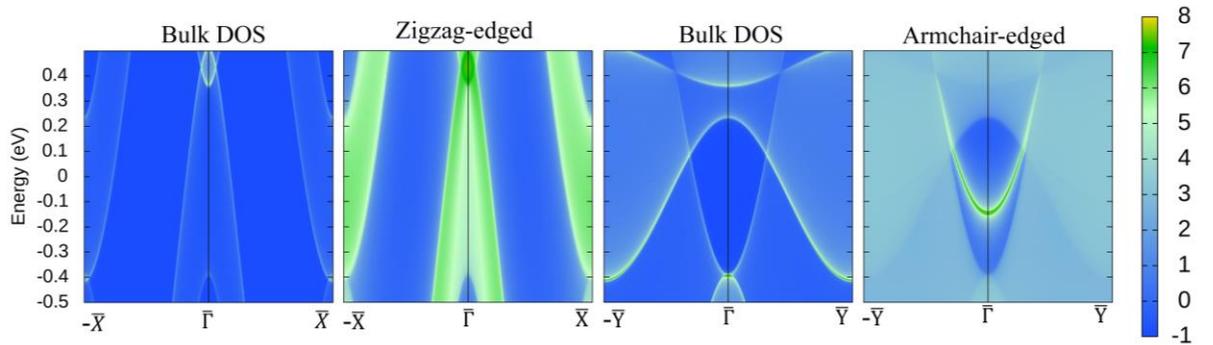

**FIG. S8.** Bulk DOS, edge DOS along the zigzag and armchair cut (see Fig. 1) for the

C$_4$BN.

**Table S1.** Structural information of the predicted stable $C_{6-2x}(BN)_x$ BPN systems.

| Phases | Lattice Parameters (Å, °) | Atoms | Fractional coordinates | | |
|---|---|---|---|---|---|
| | | | x | y | z |
| *P2/m* | a = 3.8644 | C | 0.5080 | 0.1804 | 0.5000 |
| $C_2(BN)_2$ | b = 4.5281 | C | 0.5043 | 0.8727 | 0.5000 |
| | c = 20.000 | B | 0.8251 | 0.6782 | 0.5000 |
| | $\alpha$ = 90.0 | B | 0.1872 | 0.3750 | 0.5000 |
| | $\beta$ = 90.0 | N | 0.2070 | 0.6976 | 0.5000 |
| | $\gamma$ = 90.0 | N | 0.8053 | 0.3556 | 0.5000 |
| *Pmm2* | a = 3.7945 | N | 0.5061 | 0.1975 | 0.5000 |
| $C_4BN$ | b = 4.5986 | C | 0.8169 | 0.6798 | 0.5000 |
| | c = 20.000 | C | 0.1954 | 0.6798 | 0.5000 |
| | $\alpha$ = 90.0 | C | 0.8102 | 0.3646 | 0.5000 |
| | $\beta$ = 90.0 | C | 0.2020 | 0.3646 | 0.5000 |
| | $\gamma$ = 90.0 | B | 0.5061 | 0.8731 | 0.5000 |

**Table S2. Parameters used for the band structures from Fig. 7.**

| | (a) | (b) | (c) | (d) | (e) |
|---|---|---|---|---|---|
| $\epsilon_C$ | -0.95 | -0.95 | -1.50 | -1.80 | -2.00 |
| $\epsilon_B$ | 0 | 0 | 1.00 | 1.75 | 2.10 |
| $\epsilon_N$ | 0 | 0 | -1.00 | -1.25 | -1.50 |
| $t_\alpha$ | 2.59 | 2.59 | 2.59 | 2.59 | 2.59 |
| $t_\beta$ | 2.59 | 2.59 | 2.30 | 2.15 | 2.00 |
| $t_\gamma$ | 2.59 | 2.59 | 2.40 | 2.35 | 2.30 |
| $t_\delta$ | 0.45 | 0.95 | 0.95 | 1.30 | 1.50 |
| $t_\varepsilon$ | 0.45 | -0.10 | -0.10 | -0.15 | -0.20 |
| $t_\epsilon$ | 0 | 0 | 0.50 | 0.80 | 0.90 |
| $t_\zeta$ | 0 | 0 | -0.20 | -0.25 | -0.30 |